\documentclass[12pt,a4paper]{article}
\usepackage{graphicx}% Include figure files
\usepackage{dcolumn}% Align table columns on decimal point
\usepackage{bm}% bold math
\usepackage{authblk}

\begin{document}

%\preprint{APS/123-QED}

\title{Topological properties of bound-states-in-the-continuum in geometries with broken anisotropy-symmetry}% Force line breaks with \\

\author[1]{Samyobrata Mukherjee}
%\affiliation{ICFO-Institut de Ci\`{e}ncies Fot\`{o}niques, The Barcelona Institute of Science and Technology,  08860 Castelldefels (Barcelona),  Spain.
%}
\author[1]{Jordi Gomis-Bresco}
%\affiliation{ICFO-Institut de Ci\`{e}ncies Fot\`{o}niques, The Barcelona Institute of Science and Technology,  08860 Castelldefels (Barcelona),  Spain.
%}
%\affiliation{Department of Applied Physics, Universitat Polit\`{e}cnica de Catalunya, Barcelona, Spain}
\author[1]{Pilar Pujol-Closa}
%\affiliation{ICFO-Institut de Ci\`{e}ncies Fot\`{o}niques, The Barcelona Institute of Science and Technology,  08860 Castelldefels (Barcelona),  Spain.
%}
\author[1,2,*]{David Artigas}
 %\email{david.artigas@icfo.eu}
 
\author[1,2]{Lluis Torner}
%\affiliation{ICFO-Institut de Ci\`{e}ncies Fot\`{o}niques, The Barcelona Institute of Science and Technology,  08860 Castelldefels (Barcelona),  Spain.
%}
%\affiliation{Department of Signal Theory and Communications, Universitat Polit\`{e}cnica de Catalunya,  08034 Barcelona, Spain
%}%

\affil[1]{ICFO-Institut de Ci\`{e}ncies Fot\`{o}niques, The Barcelona Institute of Science and Technology,  08860 Castelldefels (Barcelona),  Spain.}
\affil[2]{Department of Signal Theory and Communications, Universitat Polit\`{e}cnica de Catalunya,  08034 Barcelona, Spain}
\affil[*]{david.artigas@icfo.eu}

\date{}% It is always \today, today,
\maketitle
\begin{abstract}
Waveguiding structures made of anisotropic media support bound states in the continuum (BICs) that arise when the radiation channel of otherwise semi-leaky modes is suppressed. Hitherto, only structures with optical axes aligned in symmetric orientations inside the waveguide plane, where BICs appear as lines in the momentum-frequency dispersion diagram, have been considered. Here we address settings where such symmetry is broken and unveil a number of fundamental new features. Weak and strong symmetry-breaking regimes are identified, corresponding to azimuthal and polar optical axes orientation asymmetries, respectively. The azimuthal symmetry-breaking is found to still preserve the existence loci of BICs in the momentum-frequency dispersion diagram as lines. However, all possible BICs become interferometric, while the polarization separable states that occur in symmetric settings cease to exist. The polar symmetry-breaking has stronger effects and transforms the BICs' existence loci from lines to points, which correspond to full-vector states that exist at discrete values of the optical axis orientation for a given wavelength. Such transformation results in fundamental changes in the topological properties of the radiated field around the BICs. 

\end{abstract}

%\keywords{BICs, Topological Charge, Anisotropy}%Use showkeys class option if keyword

\section{\label{sec:level1}Introduction}
\noindent
Bound states in the continuum are resonant states that are radiationless despite having energies above the confining potential so that they coexist with the continuum of radiating waves. The concept was introduced by Von Neumann and Wigner in the early days of quantum mechanics \cite{Neuman1929} and revisited by Stillinger and Herrick \cite{Stillinger1975}. More recently, BICs have been shown to be a general wave phenomenon relevant to many areas of physics and technology. They were first observed in acoustic systems \cite{Parker1966} and then subsequently studied in several photonic systems, which initially included dielectric gratings \cite{Marinica2008} and waveguides with defects \cite{Bulgakov2008}. Landmark experimental demonstrations conducted in waveguide arrays \cite{Plotnik2011,Corrielli2013} and in photonic crystal slabs \cite{Hsu2013} exposed the potential of the concept to generate both fundamental new phenomena and practical consequences. BICs have been studied in a wide variety of photonic systems including diffraction gratings \cite{Monticone2017}, near zero refractive index materials \cite{Li2017}, and various periodic structures \cite{Bulgakov2017a,Bulgakov2014,Bulgakov2015}. BICs have promising applications in acoustics \cite{Linton2007}, spin filters \cite{Vallejo2010} and spin polarizers \cite{Ramos2014} for spintronics, three-dimensional photonic nanostructures \cite{Monticone2014}, resonant elements  \cite{Yoon2015}, high-Q supercavity modes \cite{Rybin2017}, embedded detectors \cite{Gansch2016} and laser resonators \cite{Kodigala2017,Midya2018,Kartashov2017}, among others that, in general, seek to improve integrated photonic circuitry. \\
\indent In general, BICs can be classified into two categories, according to their underlying existence mechanism, namely, symmetry/separability protection or interference through parameter tuning \cite{Hsu2016}. The former may disappear under the effect of perturbations, while the latter, though in some cases topologically protected, requires suitable symmetric layouts that, when broken, cause losses to appear, hindering the device performance due to the sensitivity of BICs to perturbations of symmetry. This is, for example, the case of BICs arising from wave equation separability \cite{Rivera2016} when the symmetry of the corresponding potentials is broken. Similarly, BICs existing in photonic crystal slabs \cite{Hu2018} have been shown to couple to the continuum and hence become finite resonances when the supporting structures become asymmetric.\\
\indent Recently \cite{Gomis-Bresco2017}, anisotropy-induced BICs were identified as a new class of states existing in waveguides containing anisotropic materials. Radiative-loss minima of leaky modes of planar waveguides and spectrally-embedded resonances in layered media, which actually arise from BICs, had been studied theoretically previously \cite{Marcuse1979,Shipman2013}. In general, anisotropic waveguides sustain both totally-guided and semi-leaky full-vector hybrid modes \cite{Marcuse1979,Knoesen1988,Torner1993}, and cannot be reduced to a scalar approach. The semi-leaky modes are mathematical constructions that approximately describe fields that are partially localized around the waveguide but that leak energy that is carried away towards the waveguide substrate by a continuous band of radiation modes \cite{Hu2009,Monticone2015}. They are improper modes that feature a complex eigenvalue whose imaginary part, however, gives a good approximation of the actual radiation losses when radiation is not too high. Unlike other photonic systems, such as traditional photonic crystal slabs \cite{Hsu2013}, which have radiation channels associated with transverse electric or magnetic (TE/TM) polarizations, in the anisotropy-induced BICs, radiation leakage occurs for one of the leaky mode orthogonal polarizations. When the radiation channel is cancelled by destructive interference, or when the mode features a polarization orthogonal to the radiation channel, interferometric and polarization separable BICs can exist, respectively. Polarization separable BICs are pure TE or TM modes propagating in a direction exactly orthogonal to the optical axis orientation of the crystals, while interferometric BICs feature full-vector hybridity and tunable angular propagation directions. BICs in structures built with anisotropic materials have recently been studied as defect modes with tunable Q-factors \cite{Timofeev2018}.\\
\indent Although, by and large, the anisotropic structures analyzed in Ref \cite{Gomis-Bresco2017} may be asymmetric in terms of refractive index values, the optical axes of all involved crystals were arranged in exactly anisotropy-symmetric geometries, i.e., they were aligned with each other and, importantly, laid fully within the waveguide plane. Here, we demonstrate that breaking such anisotropy-symmetry introduces fundamentally new physical properties. In particular, we find that in structures where the optical axes of the materials lay in the waveguide plane but are not aligned with each other, so that the azimuthal anisotropy-symmetry is broken, the loci of the existing BICs are lines contained within the sheet of leaky modes in the frequency-momentum dispersion diagram, similar to what was found in structures with symmetric anisotropy. However, polarization separable-BICs can no longer be supported and only interferometric-BICs with hybrid polarization exist. An even stronger transformation occurs in structures where the polar anisotropy-symmetry is broken, to the extent that the loci of existing BICs are reduced to points. Consequently, while a variation in the carrier wavelength only displaces the BIC to a different propagation direction in an on-plane symmetric or in azimuthally symmetry-broken structures, structures with polar symmetry-breaking can support BICs only for specific values of both the optical axis orientation and the wavelength. Such BICs correspond to phase singularities in the dispersion diagram that undergo a topological transition from phase jumps to screw phase dislocations together with the transformation from lines to points in the dispersion diagram. The corresponding winding numbers in the parameter space of optical axis orientations appear or disappear in such a way that the total topological charge is preserved.\\

\section{\label{sec:level1}Theory}
\noindent
We address the class of waveguides illustrated by the structure depicted in Figure 1a. $X$ is the axis normal to all the interfaces and $Y$ the wave propagation direction. In the calculations, the film thickness $d$ is conveniently normalized to the wavelength of light $\lambda$. The refractive index of the isotropic cladding and the ordinary and extraordinary indices for the substrate and the film are given by $n_c$, $n_{os}$, $n_{es}$, $n_{of}$, and $n_{ef}$, respectively. When the optical axis is oriented along the $X$ direction, the corresponding permittivity tensor is diagonal and given by:
\begin{equation}
	\hat{\epsilon}=\left[ { \begin{array}{cccc}
    n^{2}_{e} & 0 & 0\\
    0 & n^{2}_{o} & 0\\
    0 & 0 & n^{2}_{o}
    \end{array} } \right].
\end{equation}
Different optical axis orientations in the substrate $\left\lbrace\theta_s, \phi_s\right\rbrace$ and the film $\left\lbrace\theta_f, \phi_f\right\rbrace$ transform (1) by the appropriate rotation matrices $R_x(\phi)$ and $R_z(\theta)$. Guided and leaky modes in arbitrarily layered structures can be found using Berreman's transfer-matrix formalism \cite{Berreman1972}. The dimensionless eigenvalues corresponding to the ordinary and extraordinary waves, $\kappa_o$ and $\kappa_e$, are given by
\begin{equation}
\kappa_o=\pm\sqrt{\epsilon_{o} - \kappa_{y}^{2}},
\end{equation}
\begin{equation}
\kappa_e=\frac{1}{\epsilon_{xx}} \left[- \epsilon_{xy} \kappa_{y} \pm \sqrt{\epsilon_{o} \left(\epsilon_{xx} \epsilon_{e} + \kappa_{y}^{2} \left(\epsilon_{zz} - \epsilon_{e} - \epsilon_{o}\right)\right)}\right].
\end{equation}
Here $\kappa_y$ is the normalized propagation constant (effective index) in the $Y$-direction; $\epsilon_e$ and $\epsilon_o$ denote the extraordinary and ordinary permittivity, respectively, and $\epsilon_{ij}$ denotes the values of the permittivity tensor for an arbitrary orientation of the optical axes. The eigenvectors for the ordinary and extraordinary eigenvalues are respectively given by\\
\begin{equation}
\overrightarrow{F}_o=\left[\begin{matrix}\kappa_{o} \sin{\left (\phi \right )} \sin{\left (\theta \right )}\\\epsilon_{o} \sin{\left (\phi \right )} \sin{\left (\theta \right )}\\- \kappa_{o} \sin{\left (\theta \right )} \cos{\left (\phi \right )} + \kappa_{y} \cos{\left (\theta \right )}\\\kappa_{o} \left(\kappa_{o} \sin{\left (\theta \right )} \cos{\left (\phi \right )} - \kappa_{y} \cos{\left (\theta \right )}\right)\end{matrix}\right],
\end{equation}
and
\begin{equation}
\overrightarrow{F}_e=\left[\begin{matrix}- \kappa_{e} \kappa_{y} \cos{\left (\theta \right )} + \kappa_{o}^{2} \sin{\left (\theta \right )} \cos{\left (\phi \right )}\\\epsilon_{o} \left(\kappa_{e} \sin{\left (\theta \right )} \cos{\left (\phi \right )} - \kappa_{y} \cos{\left (\theta \right )}\right)\\\epsilon_{o} \sin{\left (\phi \right )} \sin{\left (\theta \right )}\\- \epsilon_{o} \kappa_{e} \sin{\left (\phi \right )} \sin{\left (\theta \right )}\end{matrix}\right],
\end{equation}
where the four rows of $\overrightarrow{F}_e$ and $\overrightarrow{F}_o$ correspond to the field components $E_y$, $z_0H_z$, $E_z$ and $z_0H_y$, respectively, with $z_0$ being the vacuum impedance. In the transfer matrix formalism, the field matrix $\hat{F}$ is defined by arranging the eigenvectors in four columns sorted as: ordinary forward ($\kappa_o^+$), ordinary backward ($\kappa_o^-$), extraordinary forward ($\kappa_e^+$), and extraordinary backward ($\kappa_e^-$), corresponding to the four eigenvalues in equations (2) and (3), where forward and backward refers to the X direction. The total field at a point in the layer is given by $\vec{m}=\hat{F}\vec{a}$, where $\vec{a}$ is a column vector that contains the coefficients for each of the basis waves. The field coefficients in $\hat{F}$ can be transformed within two points in the same layer separated by a distance $d$ using the phase matrix
\begin{equation}
\hat{A}_d={diag}\left(e^{-ik^+_od};e^{-ik^-_od};e^{-ik^+_ed};e^{-ik^-_ed}\right),
\end{equation}
with the transversal momentum being $k_i^\pm=2\pi\kappa_i^\pm/\lambda_0$. Thus, the characteristic matrix $\hat{M}$ of a film of thickness $d$ can be written using the phase matrix and the field matrix as \cite{McCall2015}
\begin{equation}
\hat{M}=\hat{F}_f^{-1}\hat{A}_d\hat{F}_f,
\end{equation}
where the subscript $f$ refers to the film parameters. For guided modes, the outgoing waves in both media are selected, while the incoming ones are set to zero. This prescription for sorting the eigenvalues and eigenvectors into forward/backward, ordinary/extraordinary waves breaks down when solving for leaky modes with complex propagation constants. Then, the incoming wave in the radiative channel must be selected properly to obtain the correct leaky mode solution, because leaky modes are improper solutions that do not evanescently vanish at $x\to-\infty$. Once we have selected the four waves in the cladding and substrate, we use the film characteristic matrix $\hat{M}$ to apply the boundary conditions. Assuming that the nonzero coefficients of the basis vector for the ordinary and extraordinary wave are $a_c^{TE}$ and $a_c^{TM}$ at the cladding and $a_s^o$ and $a_s^e$ at the substrate, and using boundary conditions, the fields at the substrate-film interface fulfill:
\begin{equation}
	a^o_s\cdot\vec{v}^o_s + a^e_s\cdot\vec{v}^e_s = a_c^{TE}\cdot\hat{M}\vec{v}_c^{TE} + a_c^{TM}\cdot\hat{M}\vec{v}_c^{TM},
\end{equation}
where $\vec{v}_j^i$ is a $4\times 1$ column vector that contains the field components of the basis wave (ordinary, extraordinary, TE and TM at the superscript) in the substrate and cladding (indicated at the subscript). Rewriting equation (8) in a matrix form, with the unknown coefficient vector $\vec{a}_T=(a_c^{TE},a_c^{TM},a_s^o,a_s^e)$, one obtains the homogeneous system of equations $\hat{W}\vec{a}_T=0$. The solution for $\kappa_y$ is thus given by solving $\left|⁡\hat{W} \right|=0$, which can explicitly be written as:
\begin{equation}
	\left| {\begin{array}{cccc}
   -t_{c,E_y}^{TE} & -t_{c,E_y}^{TM} & v_{s,E_y}^{o} & v_{s,E_y}^{e}\\
   -t_{c,H_z}^{TE} & -t_{c,H_z}^{TM} & v_{s,H_z}^{o} & v_{s,H_z}^{e}\\
   -t_{c,E_z}^{TE} & -t_{c,E_z}^{TM} & v_{s,E_z}^{o} & v_{s,E_z}^{e}\\
   -t_{c,H_y}^{TE} & -t_{c,H_y}^{TM} & v_{s,H_y}^{o} & v_{s,H_y}^{e}\\
  \end{array} } \right|=0,
\end{equation}
where $t_c^{TE/TM}=\hat{M}\vec{v}_c^{TE/TM}$. Each element of $\hat{W}$ is a function of $\kappa_y$ as seen in the components of the field expressions in equations (4) and (5). When a non-trivial solution exists for this system of homogeneous equations, the corresponding values of $\kappa_y$ denote modes of the system. Therefore, solving equation (9) for $\kappa_y$ gives us the mode indices. Both guided and semi-leaky mode indices can be obtained by choosing the appropriate waves and boundary conditions.\\
\indent BICs exist when the radiative channel is cancelled, so that $a_s^o=0\ (a_s^e=0)$ in the left-hand-side of (8) for a negative (positive) substrate with an ordinary (extraordinary) radiation channel. Then, (8) reduces to
\begin{equation}
	a_s^i\cdot\vec{v}_s^{i} -(a_c^{TE}\cdot\hat{M}\vec{v}_c^{TE}+a_c^{TM}\cdot\hat{M}\vec{v}_c^{TM})=0,
\end{equation}
where the superscript $i=e$ or $o$ stands for an ordinary or an extraordinary radiation channel, respectively. Following the same superscript notation and by using a reduced coefficient vector $\vec{a}_r^i=(a_c^{TE},a_c^{TM},a_s^i)$, equation (10) can be written in matrix form as $\hat{Z}_{ri}\vec{a}_r^i=0$, where $\hat{Z}_{ri}$ is a 4x3 matrix. The system of equations has a non-trivial solution when 
\begin{equation}
	\det(\hat{Z}_{ri})=0,
\end{equation}
where the determinant is calculated for any combination of rows of $\hat{Z}_{ri}$. Equation (11) is of particular importance, as it provides the condition for which a leaky mode becomes a BIC. Together with equation (9), equation (11) makes it possible to obtain the existence loci of BICs \cite{Gomis-Bresco2017}.\\

\section{Results and discussion}
\noindent As we describe above, the propagation direction is kept in the Y direction. Then, when a sample rotates with respect to the X axis, the azimuthal angles for the OAs in the film and substrate, $\phi_f$ and $\phi_s$, rotate concurrently. Therefore, we chose to use as working parameters the azimuthal orientation of the film OA, $\phi_f$, and define the substrate azimuthal angle through the detuning $\Delta\phi=\phi_s-\phi_f$. Note that if one takes the film OA projection at the interface of the structure as a reference, then $\phi_f$ specifies the propagation direction instead of $k_z$ and $k_y$ ($\phi_f=\arctan⁡{(k_z /k_y)}$), and $\Delta\phi$ can be used as a fixed parameter. Thus, the dispersion diagrams shown below depict projections of the 3D momentum-frequency dispersion diagram to a 2D representation in the $d/\lambda-\phi_f$ plane, and as in the 3D plot of Fig. 1b, the color represents the decay length. In what follows, BICs are studied using the analysis described above and keeping the refractive indices constant.\\
\begin{figure*}
\begin{center}
\includegraphics[width=.7\textwidth]{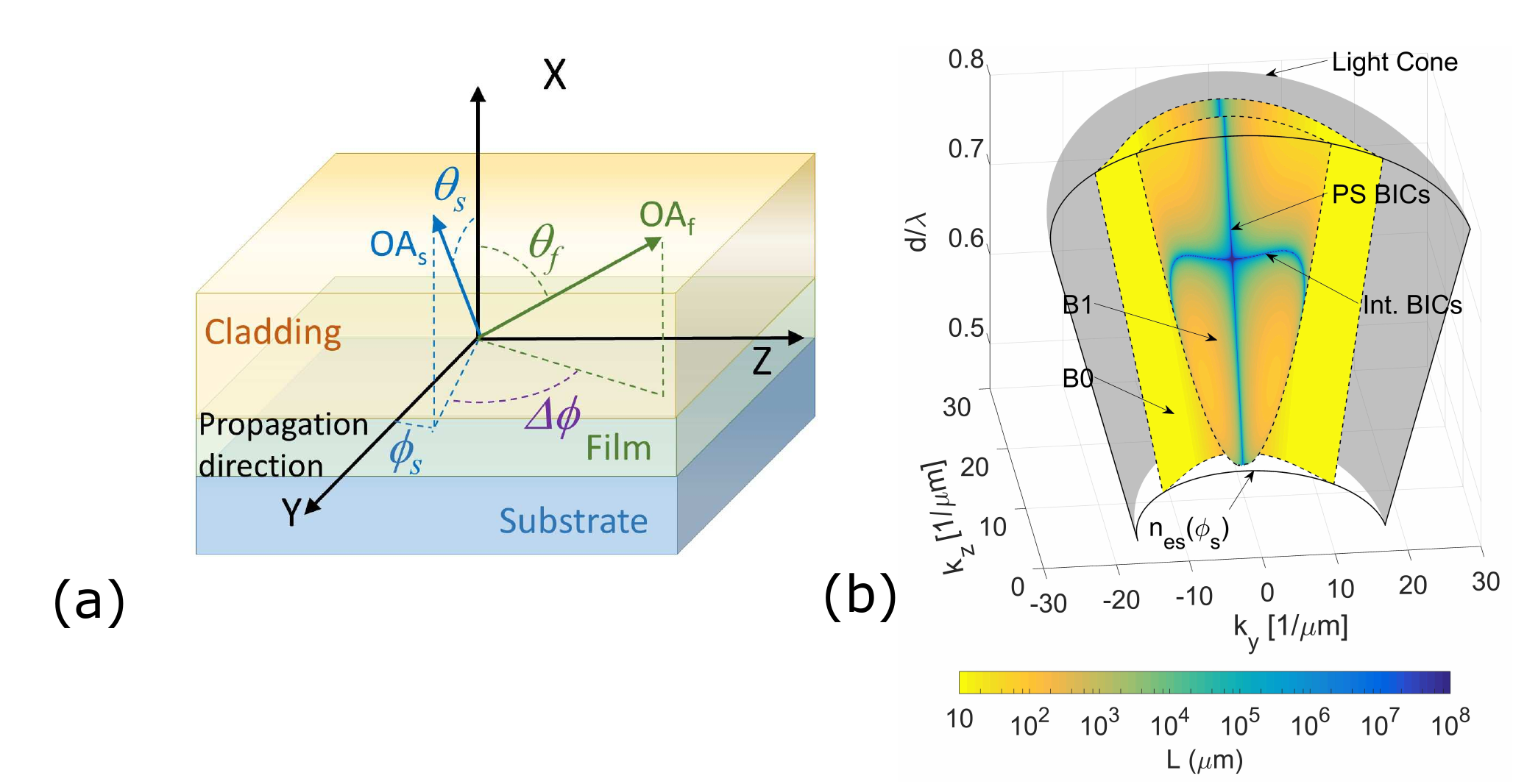}
\caption{a) Layout of the System: A generic layered waveguide system comprising cladding, film and substrate. Light propagates along the $Y$ direction. The green and the blue arrows indicate the substrate and film optical axis (OA), respectively. The polar $\theta_f$, and azimuthal $\phi_f$ angles indicate the film OA orientation in spherical coordinates. The substrate OA orientation is determined by $\theta_s$ and the azimuthal detuning by $\Delta\phi=\phi_s-\phi_f$. b) Momentum-frequency dispersion diagram for the leaky modes of an anisotropy-symmetric structure with a film with positive birefringence ($n_{of}=1.5$, $n_{ef}=1.75$), a substrate with a negative birefringence ($n_{os}=2.00$, $n_{es}=1.25$) and air as a cladding $n_{c}=1$. The polar orientation of the OA are $\theta_s=\theta_f=90^{\circ}$. The ratio $d/\lambda$ is the normalized frequency and $k_z$ and $k_y$ are the mode momenta along the crystallographic axis (with the OA laying in the Z direction), which can be related to the angle $\phi_f$ between the OA orientation and the propagation direction as $\phi_f=\arctan⁡{(k_z /k_y)}$. Two yellowish surfaces (B0 and B1) for the leaky modes are plotted, where dashed lines show the limits of the surface and the color indicates the decay length $L$ (in $\mu m$). Then, blue lines stand for an infinite decay length, indicating the existence of BICs. The labels INT-BICs and PS-BICs refer to interferometric and polarization-separable BICs, respectively. The grey surface behind the leaky mode sheets is the light cone, given by $k_0\cdot n_{os}$; the light grey-transparent surface limited by solid lines sets the leaky mode cutoff, given by $k_0\cdot n_{es} (\phi_f)$. Here $k_0$ is the free-space wavenumber. 
\label{f1}}
\end{center}
\end{figure*}
Figure 1b shows the momentum-frequency dispersion diagram ($k_y$, $k_z$, $d/\lambda$) of an illustrative anti-guiding structure with anisotropy-symmetry that does not support standard guided modes. However, it supports two families of leaky modes, which are shown in the plot with a color proportional to their decay length. For a fixed wavelength, the families reduce to two branches of leaky modes. The radiation channel is related to the polarization at which the substrate refractive index is higher than the effective index of the leaky mode, while the orthogonal polarization is guided. BICs are pure real solutions at which the leakage through the radiation channel vanishes. The solution for the BICs exist as lines in the dispersion diagram, in contrast to what happens in photonic systems with degenerated radiation channels, where they exist as dots. This can be seen in Fig. 1b, where BICs appears as blue lines. The vertical line and the curved horizontal line in the upper surface correspond to polarization separable BICs (the polarization is orthogonal to the radiation channel) and interferometric BICs (the radiating ordinary wave is cancelled by destructive interference), respectively. The lower surface corresponds to the fundamental leaky mode branch, and only supports polarization separable BICs. Note that such BICs are the \textit{only non-radiating states} of the structure.\\
\indent Figure 2 shows the projection of the dispersion diagram for the second branch of leaky modes (labelled in Fig. 1b as B1). Figure 2a corresponds to an anisotropy-symmetric structure (all optical axes are fully in-plane and perfectly aligned with each other) and the BIC existence lines are therefore symmetrically distributed with respect to $\phi_f=90^{\circ}$ in the dispersion diagram. Polarization separable BICs (central vertical line) carry a pure TE polarization, while the interferometric BICs (curved lateral lines) feature a hybrid (TE-dominant) polarization. The anisotropy-symmetry is broken when the azimuthal OA orientations in the film and substrate are shifted relative to each other (i.e., $\Delta\phi=\phi_s-\phi_f\neq 0^{\circ}$), or when the OAs are oriented out of the plane (i.e., $\theta_f$ and/or  $\theta_s\neq90^{\circ}$), and in general combinations of both geometries. Importantly, note that the rotation of the OA of the substrate or the film, either in the azimuthal or the polar direction, modifies the modes supported by the system, but also the coupling between the modes and the continuum \cite{Sadreev2017}.\\
\begin{figure*}
\begin{center}
\includegraphics[width=.7\textwidth]{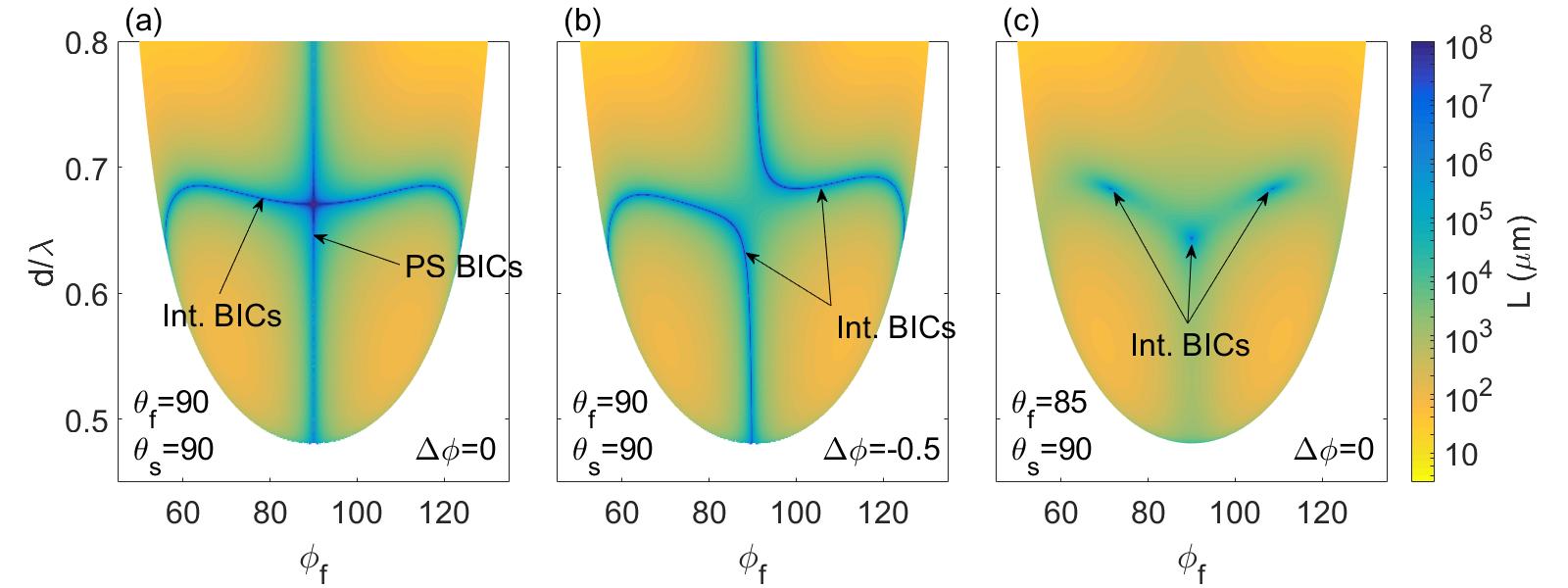}
\caption{Existence loci of BICs in structures with broken anisotropy-symmetry. The colored area in the figures shows the decay length of the leaky modes in the $\phi_f-d/\lambda$ plane, where BICs appear as blue lines. The transition from the colored to the white area is the cut-off for the leaky modes. a) Anisotropy-symmetric structure shown in Fig. 1b, where the substrate and film OAs are aligned and in-plane. b) Structure with weakly broken-anisotropy-symmetry in the azimuthal direction ($\Delta\phi=-0.5^{\circ}$). c) Structure with strongly broken-anisotropy-symmetry in the polar direction, by taking the OA of the film out of the interface plane ($\theta_s=90^{\circ}$, $\theta_f=85^{\circ}$) and keeping $\Delta\phi=0^{\circ}$ (i.e., the projection of the film OA is parallel to the substrate OA). Note the transition from existence lines in (a) and (b) to dots in (c).  
\label{f2}}
\end{center}
\end{figure*}
\indent Figure 2b shows the projection of the dispersion diagram for a structure where only the azimuthal symmetry is broken with $\Delta\phi=-0.5^{\circ}$, while the polar symmetry still holds ($\theta_f=\theta_s=90^{\circ}$). Consequently, the symmetry in the dispersion diagram with respect to $\phi_f=90^{\circ}$ axis is also broken, resulting in a distortion of the existence lines of BICs. Large values of $\Delta\phi$ result in greater distortions, thus affording a way to tune the BIC existence direction. The crossing of the polarization separable BICs and interferometric BICs in Fig. 2a transforms into an anti-crossing in Fig. 2b, where the upper/right and lower/left blue lines correspond exclusively to interferometric BICs. Therefore, pure TE or TM modes do not exist and polarization separable BICs transform into hybrid interferometric BICs. Therefore, breaking the azimuthal anisotropy-symmetry results in the smooth distortion of the BIC lines in the dispersion diagram. However, such behavior changes dramatically when the polar symmetry is broken. Fig. 2c shows the projection of the dispersion diagram for a structure where only the polar symmetry is broken because the OA in the film is out-of-plane ($\theta_f=85^{\circ}$ and $\theta_s=90^{\circ}$) while the azimuthal symmetry holds ($\Delta\phi=0^{\circ}$). In this case, the existence lines of BICs in Fig. 2a transform into dots of interferometric BICs in Fig. 2c so that there is only one combination of a specific orientation of the OA of the film $\phi_f$ and normalized wavelength $d/\lambda$ that results in a BIC. Thus, adding a new degree of freedom in the parameter space reduces the dimensionality of the BIC existence loci, so that continuous lines transform into discrete dots in the $\phi_f-d/\lambda$ dispersion diagram.\\
\indent To gain deeper insight into the difference between breaking the polar and azimuthal anisotropy-symmetry, the full vector character of the leaky modes must be taken into account. In the particular case shown, since $\theta_f=85^{\circ}$, the orientation of the polarization of the ordinary wave at the film evolves with the angle $\phi_f$, while the orientation of the polarization of the ordinary radiation channel, which determines the continuum, remains constant because $\theta_s=90^{\circ}$. Consequently, the existence of interferometric BICs requires the fulfillment of an additional constraint, namely, the matching of the polarization, in order to produce destructive interference of the radiation channel. We found that only discrete points of the BIC existence lines in Fig. 2a fulfill this condition, as depicted in in Fig. 2c. This feature is a unique and direct consequence of the vector nature of the anisotropy-induced BICs.\\
\indent The transition from existence lines to dots in the projection of the dispersion diagram is best elucidated  by analyzing the phase of the radiation fields around a BIC. Figure 3a shows the case of a structure with anisotropy-symmetry. BIC existence lines correspond to zeros in the amplitude of the radiation channel and feature a phase dislocation in the $\phi_f-d/\lambda$ parameter space, with a jump of $\pm\pi$ (the phase is measured relative to the phase of the extraordinary wave component, which is confined). The existence lines of BICs separate regions featuring radiation fields with opposite phases (reddish and bluish color in Fig. 3a, respectively). The phase dislocations are preserved when only the azimuthal anisotropy-symmetry is broken (Fig. 3b).\\
\begin{figure*}
\begin{center}
\includegraphics[width=.7\textwidth]{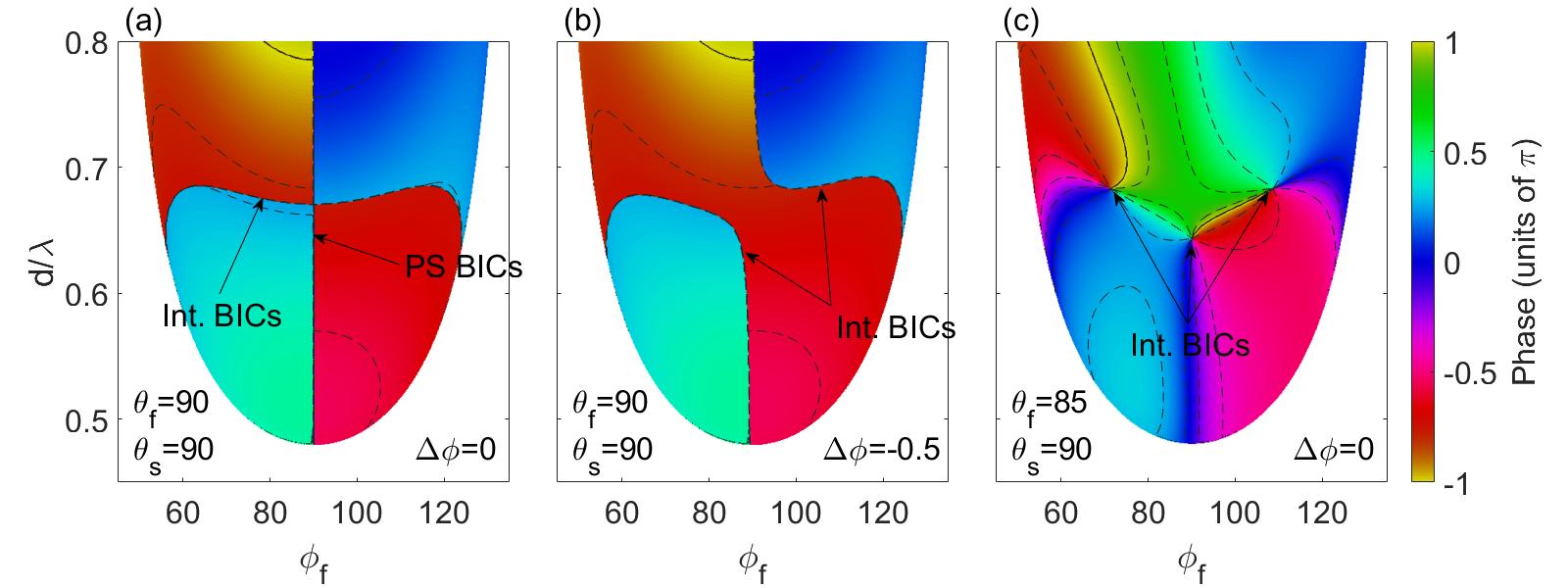}
\caption{Transformation from simple phase dislocation to screw phase dislocations. The parameters are the same as in Fig. 2. The color stands for the phase difference of the ordinary radiative channel with respect to the phase of the extraordinary confined wave in the substrate. The existence lines of BICs in (a) and (b) exhibit phase dislocation of $\pi$. BICs in (c) are singularities with winding numbers $+1$ (clockwise increase of phase) and $-1$ (anti-clockwise increase of phase).  
\label{f3}}
\end{center}
\end{figure*}
In contrast, when the polar anisotropy-symmetry is broken, the transition from existence lines to dots in Fig. 2c results in a topological change from a simple phase dislocation to screw phase dislocations in Fig. 3c. At orientations of the film optical axis in the range $0^{\circ}<\phi_f<180^{\circ}$, the winding numbers of the BIC points originated from the interferometric BICs and the polarization separable BICs are $-1$ and $+1$, respectively. They reverse sign for $180^{\circ}<\phi_f<360^{\circ}$ (not shown in the figure). Transforming Fig. 3 from the parameter space $\phi_f-d/\lambda$ into the momentum space $k_y-k_z$, reveals that the screw phase dislocation appears in the radiated far field which thus carries orbital angular momentum. The resulting singularity is a phenomenon similar to the appearance of vortices in other photonic structures that support BICs \cite{Zhen2014,Bulgakov2017}, however, a similar topological transition has not been described in any other physical setting.\\
\indent The topological transition in the parameter space $\phi_f-d/\lambda$ from lines featuring phase dislocations to dots featuring a screw phase is far from trivial and greatly depends on the parameters of the structure. A variety of combinations of positive and negative birefringence for the film and substrate, which lead to the corresponding diversity of standard modal spectroscopies for the structure \cite{Knoesen1988} may be considered. For example, Fig. 4 shows the phase of the radiation fields to the substrate for a structure made of a positive and a negative birefringent substrate and film, respectively. In this case, BICs for the anisotropy-symmetric structure (Fig. 4a) and for a structure with broken azimuthal anisotropy-symmetry ($\Delta\phi=-10^{\circ}$ in Fig. 4b) appear as vertical lines that feature phase dislocations of $\pi$, similar to Fig. 3a-b. However, in this structure, the breaking of the polar symmetry does not necessarily result in a topological transition from phase dislocations of $\pi$ to screw phase dislocations. This is the case in Fig. 4c, where BICs disappear for structures with broken polar anisotropy-symmetry by taking the film OA out of the interface plane. BICs exist again if, in addition to the film OA, the OA of the substrate is also taken out-of-plane. Then, when the OA of the substrate is mainly oriented towards the substrate (same direction as the film OA) an interferometric BIC with screw phase dislocation of charge +1 appears at $\phi_f=90^{\circ}$ (Fig. 4d). Alternatively, if the OA of the substrate is oriented towards the film (opposite to the film OA), two Interferometric BICs with screw phase dislocation of charge $-1$ appear (Fig. 4e). Finally, when in addition to polar symmetry, the azimuthal symmetry is broken, the loci of the BIC showing screw phase dislocation is only translated to a different point $\phi_f$ in the $\phi_f-d/\lambda$ space (Fig. 4f), without modifying its topological nature.\\ 
\begin{figure*}
\begin{center}
\includegraphics[width=.7\textwidth]{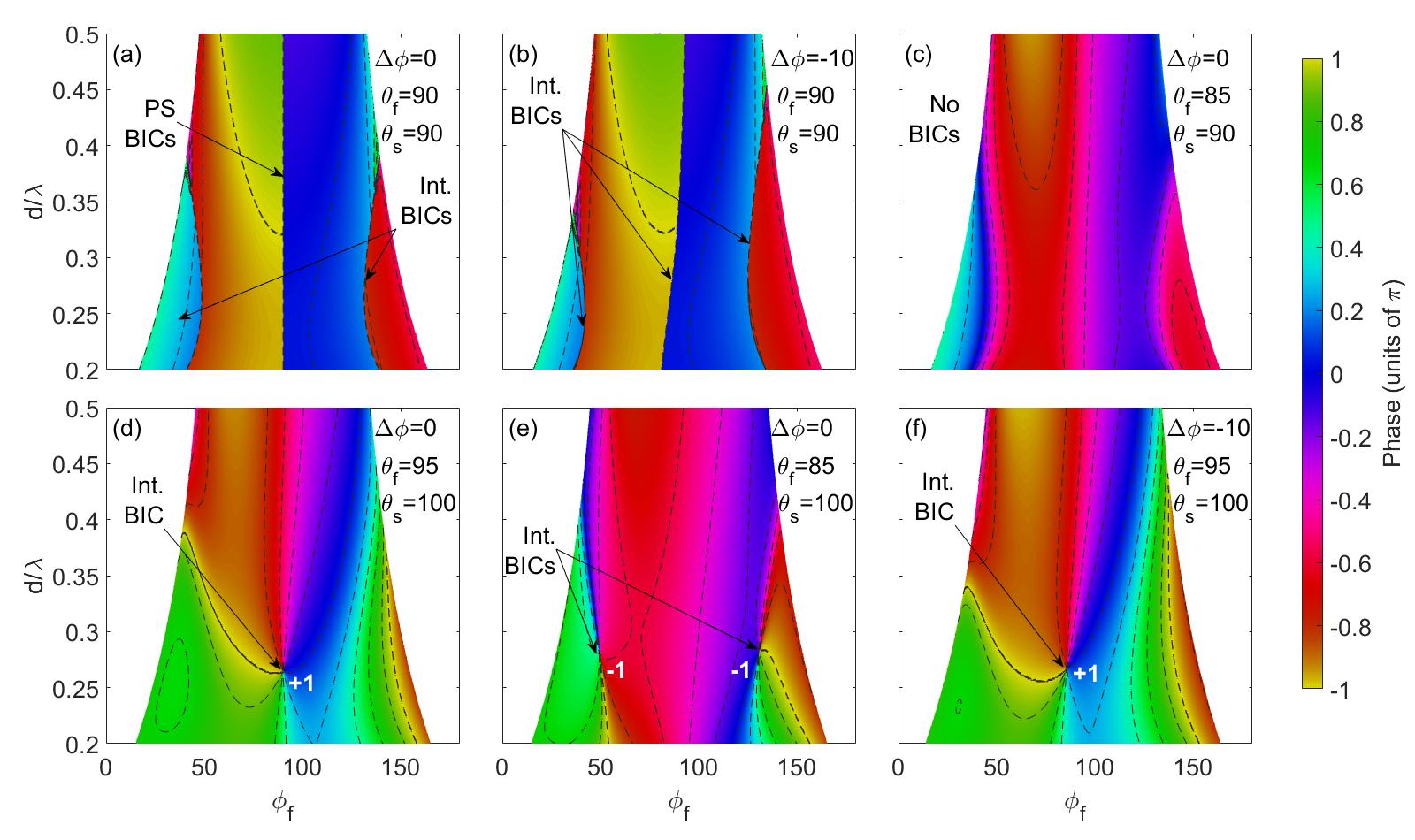}
\caption{Topological transformation of the phase of the radiative channel in a structure with a positive birefringent substrate and negative film. The parameters of the structure are: $n_{es}=2.00$, $n_{os}=1.25$, $n_{ef}=1.50$, $n_{of}=1.75$, $n_c=1$, $\theta_s=\theta_f=90^{\circ}$, $\Delta\phi=0^{\circ}$. The color stands for the phase difference between the extraordinary radiative channel and the ordinary wave in the substrate. The lines of existence of BICs in an anisotropy-symmetric structure (a) and in structures with the azimuthal symmetry broken (b) exhibit phase dislocations of $\pi$. Breaking the polar symmetry by taking the film optical axis out of plane precludes the existence of BICs (c). BICs occur as points when the polar symmetry is broken by taking both, the film and substrate optical axes out of plane, resulting in one BIC with winding number $+1$ (d) or two BICs with winding number $-1$ (e) depending on the relative orientation of the film and substrate OAs. When both, polar and azimuthal symmetries are broken, the BIC loci can be tuned for any orientation of $\phi_f$ (f).  
\label{f4}}
\end{center}
\end{figure*}
\indent The above results suggest that there are two different symmetry-breaking regimes, which we refer to as weak and strong, that correspond to the breaking of the azimuthal and the polar anisotropy-symmetries, respectively. The former is characterized by a non-vanishing $\Delta\phi=\phi_f-\phi_s$, while the latter occurs whenever the OAs are out of plane ($\theta_f$ or $\theta_s\neq90^{\circ}$). Note that here weak and strong do not refer to whether the symmetry is broken by a small or a large angle, but rather to the fact that the two mechanisms produce fundamentally different effects.\\
\indent To further explore the properties of BICs existing in both regimes, in Figure 5 we plot the orientation of the film OA,  $\theta_f$ and $\phi_f$, in a spherical representation, while $\Delta\phi=\phi_f-\phi_s$, and $\theta_s$ are kept constant. Every cut for a given value of $\theta_f$ (following a parallel) in this representation corresponds to the dispersion diagrams $\phi_f-d/\lambda$ shown in Fig. 2. In this way, in practice, each spherical representation corresponds to a structure with a fixed OA orientation for the substrate, typically a crystal, and a film made of, e.g., a liquid crystal, where the OA polar angle $\theta_f$ can be controlled by a DC electric field. Then, $\phi_f$ could be varied by the physical rotation of the sample for a fixed propagation direction.\\
\indent Figure 5a depicts the BIC existence loci, which appear as lines, and winding number (labelled as $\pm1$) for a structure with the film OA out-of-plane and the substrate OA still in the interface plane. No azimuthal asymmetry is introduced. This representation is spherically symmetric, as existence lines of BICs in the four quadrants of the sphere occur at symmetric locations with respect to the $\theta_f=\phi_f=90^{\circ}$  point, except that the winding numbers change sign with respect to the equatorial plane. The plot shown in Fig. 2a, obtained for a structure with symmetric anisotropy, corresponds to a cut of the sphere at $\theta_f=90^{\circ}$ (equator), and shows that interferometric BICs exist for a broad range of values of $\phi_f$ and wavelengths (Fig. 5a). The polarization separable BIC shown in Fig. 2a appears here as a white dot at the symmetry point $\theta_f=\phi_f=90^{\circ}$, where the white color indicates that the polarization separable BICs exist for all values of the wavelength above the mode cut-off. A cut at $\theta_f=85^{\circ}$ in Fig. 5a intersects the BIC existence loci at only three points, with different wavelengths falling in the green-blue region of the spectrum, thus yielding the three dots with screw phase dislocations depicted in Figs. 2c and 3c. The allowed wavelength for the three interferometric BICs to exist varies when the film OA is taken out of the interface plane. In addition, the loci of the BICs grow closer as the film OA moves further away from the interface plane, until the three BICs join near $\theta_f=46^{\circ}$ and $\theta_f=134^{\circ}$ in the northern and southern hemisphere, respectively. Then, two BICs with opposite winding numbers are cancelled, consistent with topological charge conservation. The cancellation of BICs featuring winding numbers with opposite signs also occurs when the interferometric BICs reach the equator. The winding numbers reverse their sign in the range $\phi_f=[180^{\circ}, 360^{\circ}]$ (i.e., in the reverse of the sphere), thus the total topological charge in the full range of orientations $\phi_f$ for each value of $\theta_f$ is always null.\\
\begin{figure*}
\begin{center}
\includegraphics[width=.7\textwidth]{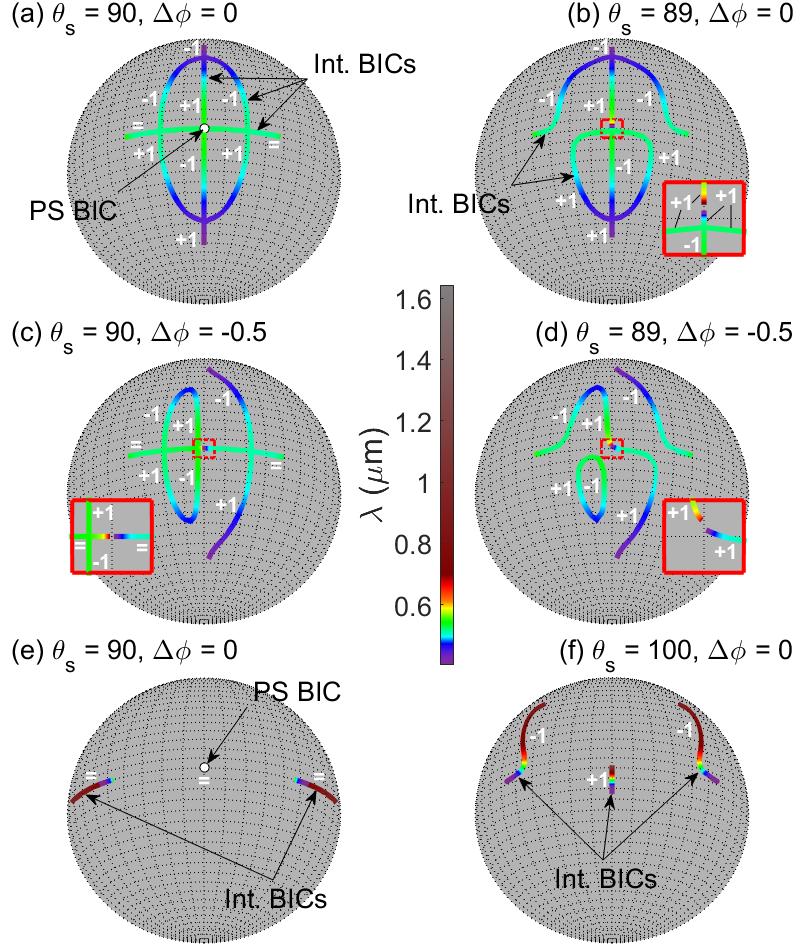}
\caption{Spherical representation of the BIC existence loci as a function of the optical axis orientation in the guiding film. The film OA orientation is given by the azimuthal angle $\phi_f$ and the polar angle $\theta_f$ in spherical coordinates, while the parameters for the substrate ($\theta_s$, and $\Delta\phi$) are kept constant in each plot. Only the hemisphere in the range $\phi_f=[0^{\circ}, 180^{\circ}]$ is depicted as the vertical axis of the sphere, $\theta_f=0^{\circ}$ is a symmetry axis. Film thickness is $d=0.35 \mu m$; hence, the colors of the BIC existence lines indicate the value of the wavelength. Winding numbers are labeled as `+1' or `-1' and phase jumps are labeled with `='. a) Spherical representation for a structure with a positive film and a negative substrate. The indices of refraction are as in Fig. 1b. The substrate OA is parallel to the interface plane ($\theta_s=90^{\circ}$, $\Delta\phi=0^{\circ}$). b) Same as in (a) but the substrate OA is out of the interface plane ($\theta_s=89^{\circ}$), resulting in strong anisotropy-symmetry breaking even at the equatorial plane. In both cases $\Delta\phi=0^{\circ}$. c) Structure with azimuthal  anisotropy-symmetry-breaking ($\Delta\phi=-0.5^{\circ}$) with the substrate OAs in the interface plane $\theta_s=90^{\circ}$. d) Combined azimuthal ($\Delta\phi=0.5^{\circ}$) and polar ($\theta_s=89^{\circ}$) anisotropy-asymmetries. The insets in (b-d) are a magnification of the area near $\theta_f=90^{\circ}$ and $\theta_f=90^{\circ}$ that show the splitting of the polarization separable BIC line in (a), shown as a white point, into two segments of interferometric BICs. e) Spherical representation of a structure with negative film and positive substrate. The indices of refraction are as in Fig. 4. The substrate OA is parallel to the interface plane ($\theta_s=90^{\circ}$, $\Delta\phi=0^{\circ}$). For this particular case, BICs cease to exist when the polar anisotropy-symmetry is broken ($\theta_f\neq90^{\circ}$). f) Same as in (e) but now the substrate OA is out of the interface plane ($\theta_s=100^{\circ}$).  
\label{f5}}
\end{center}
\end{figure*}
\indent Breaking the anisotropy-symmetry by taking the substrate OA out-of-plane results in different configurations for the northern and southern hemispheres in the spherical representation. As shown in Fig. 5b for $\theta_s=89^{\circ}$, the symmetry of the BIC existence lines in the spherical representation between the northern and southern hemispheres is broken, but the symmetry of the eastern and western hemispheres is preserved. Here, the cut at the equator no longer corresponds to a structure with anisotropy-symmetry. Then, the crossings that appear in Fig. 5a  at the equator ($\theta_f=90^{\circ}$, $\phi_s=90^{\circ}$) corresponding to interferometric BICs, now transform into anti-crossings and the lines of BIC existence are pulled out from the equator. In addition, the polarization separable BIC existence line at $\phi_s=90^{\circ}$ (shown as a white dot in Fig, 5a) transforms into two separate lines of interferometric BICs, and a gap in wavelength appears in the polar direction. The lower section of the BIC existence line crosses the equator (inset of Fig. 5b) for a given wavelength, resulting in a dot in the dispersion diagram that is the only BIC that survives at $\theta_f=90^{\circ}$ when the substrate OA is out-of-plane. When $\theta_f\neq90^{\circ}$, three isolated interferometric BICs may exist, one of them always at $\phi_f=90^{\circ}$ and two at symmetric orientations with respect to the former. Taking the substrate OA out-of-plane in the opposite direction, i.e., $\theta_s=91^{\circ}$, results in a representation where the BIC existence lines are mirror images with respect to the equatorial plane of those shown in Fig. 5b, but the sign of the winding numbers of the interferometric BICs is maintained. Again, BICs featuring opposite winding numbers cancel at crossings of BIC existence lines.\\
\indent Breaking the azimuthal anisotropy-symmetry ($\Delta\phi=0.5^{\circ}$), but keeping the substrate OA in-plane ($\theta_s=90^{\circ}$), breaks the plane mirror symmetry at the meridian plane,  $\phi_f=90^{\circ}$ and preserves the equatorial mirror symmetry between northern and southern Hemispheres (Fig. 5c). Under such conditions, the interferometric BICs existing at the equator are preserved (corresponding to Fig. 2b and 3b), but no BIC can exist at the meridian $\phi_f=90^{\circ}$. Consequently, the two crossings appearing at the top and the bottom of the interferometric BICs existence lines ($\phi_s=90^{\circ}$ and $\theta_f\neq90^{\circ}$) become anti-crossings and the polarization separable BIC existence line breaks into two existence lines of interferometric BICs (inset in Fig. 5c), opening a gap of existence in wavelength and azimuthal angle. Note that the transition to anti-crossings is a combined effect of the azimuthal and polar symmetry-breaking, as $\theta_f\neq90^{\circ}$.\\
\indent When both, azimuthal and polar anisotropy-symmetry-breaking are present with the substrate OA out of the interface plane ($\theta_s\neq90^{\circ}$), the mirror-symmetries between the north-south and the east-west hemispheres are broken and, as a consequence, we find that no crossings occur in the existence lines of BICs as illustrated in Fig. 5d. Existence lines of BICs then appear as closed lines or as disconnected lines that cease to exist at the leaky mode cut-off. The topological charge is maintained along a disconnected BIC existence line, while it switches sign at structures located at a maximum or minimum in $\theta_f$ of a closed line. If the polar symmetry-breaking increases, the closed BIC existence line may cease to exist as they collapse to a single point and thus the corresponding winding numbers cancel each other. Open BIC existence lines cease also to exist for structures where they would fall beyond the leaky mode cut-off edge, resulting in structures where no BICs exist for any orientation of the film OA.\\
\indent The spherical representation corresponding to the structure analysed in Fig. 4 is presented in Fig. 5e-f, and it shows a different topological map compared to Fig. 5a-d. The most remarkable property when the substrate OA is lying in the interface plane (Fig. 5e), is that BICs exist only at the equator, i.e., in structures where all OAs are contained in the interface plane. As a result, breaking the polar anisotropy-symmetry by taking the film OA out of the interface plane results in no BICs, as shown in Fig. 4c. When the polar symmetry is broken by taking the substrate OA out of the plane (Fig. 5f), a number of features similar to those described above occur. Namely, first, all BIC existence lines are pulled out of the equator. Then polarization separable BICs cease to exist and the resulting interferometric BICs only exist at given orientations of the film OA and wavelengths and feature screw phase dislocations in the radiated field near the BIC. Second, the winding number of the BIC originating from the polarization separable BIC ($\phi_f=90^{\circ}$) exhibits a sign opposite to the winding number corresponding to the BICs originating from interferometric BICs. The sign of the winding number changes sign in the reverse of the sphere, $\phi_f=[180^{\circ}, 360^{\circ}]$, so that the total topological charge in the full range of orientations $\phi_f$ for each value of $\theta_f$ is null. Third, the only possible BIC when the film OA is located on plane is a reminiscence of the polarization separable BIC existing at $\phi_f=90^{\circ}$. Fourth (not shown), breaking the azimuthal symmetry only deforms the BIC existence lines in the west-east direction, avoiding the existence of polarization separable BICs, but preserving the phase nature of the radiation field around a BIC, as phase jumps (as discussed in Fig. 4b) or screw phase dislocations (Fig. 4f).

\section{Conclusions}
\noindent
In closing, we stress that the fundamental ingredient of the phenomena uncovered here is the breaking of the very optical axes anisotropy-symmetry of the structures, rather than a material (e.g., refractive index) or geometrical asymmetry. Here we addressed uniaxial media and a relatively simple structure, but results are relevant to all types of natural or artificial anisotropic materials, including biaxial media, and more complex geometries where anisotropy-induced BICs may exist. We found that breaking the azimuthal anisotropy-symmetry results in weak transformations of the BIC properties, while polar symmetry-breaking causes strong changes. In particular, in the latter case, we found that the corresponding structures can support BICs only for a single, discrete combination of wavelength or carrier frequency and optical axis orientation. Also, the dispersion diagrams and radiation far-fields around the BICs exhibit a much richer topological structure than their counterparts in anisotropy-symmetric media. Our findings connect the areas of full-vector bound-states and scalar topological photonics, providing insight to the program that aims at expanding the BIC concept to general anisotropic media, a research area that remains essentially unexplored.\\
%\section{\label{sec:level1} METHODS}

\section*{Acknowledgments}
The authors acknowledge financial support of the Generalitat de Catalunya through the CERCA Programme; the Spanish Ministry of Economy and Competitiveness (MINECO) through the ‘Severo Ochoa 2016-2019' grant SEV-2015-0522 and the grant FIS2015-71559-P; the Fundació Cellex and the Fundació Mir-Puig. This project has received funding from the European Union's Horizon 2020 research and innovation programme under the Marie Sk\l odowska-Curie grant agreement No 665884.

\bibliographystyle{unsrt}

\end{document}